\documentclass{aa}  

\usepackage{natbib,twoopt}
\bibpunct{(}{)}{;}{a}{}{,}          
\makeatletter
  \newcommandtwoopt{\citeads}[3][][]{\href{http://adsabs.harvard.edu/abs/#3}
    {\def\hyper@linkstart##1##2{}
     \let\hyper@linkend\@empty\citealp[#1][#2]{#3}}}
  \newcommandtwoopt{\citepads}[3][][]{\href{http://adsabs.harvard.edu/abs/#3}
    {\def\hyper@linkstart##1##2{}
     \let\hyper@linkend\@empty\citep[#1][#2]{#3}}}
  \newcommandtwoopt{\citetads}[3][][]{\href{http://adsabs.harvard.edu/abs/#3}
    {\def\hyper@linkstart##1##2{}
     \let\hyper@linkend\@empty\citet[#1][#2]{#3}}}
  \newcommandtwoopt{\citeyearads}[3][][]
    {\href{http://adsabs.harvard.edu/abs/#3}
    {\def\hyper@linkstart##1##2{}
     \let\hyper@linkend\@empty\citeyear[#1][#2]{#3}}}
\makeatother

\usepackage{graphicx}

\usepackage{txfonts}
\usepackage{xcolor}

\usepackage{placeins}
\usepackage{lastpage}
\usepackage{subfig,float}

\begin{document} 

   \title{Connections between the cycle-to-cycle light curve and O$-$C variations of the non-Blazhko
   RR Lyrae stars}
   \titlerunning{C2C and O$-$C variations on RR Lyrae stars}

   \author{J. M. Benk\H{o}\inst{1},
          A. B\'odi\inst{2},
          E. Plachy\inst{1,3}
          \and
          L. Moln\'ar\inst{1,3}
          }

   \institute{Konkoly Observatory, HUN-REN Research Centre for Astronomy and Earth Sciences,
              MTA Centre of Excellence, \\ Konkoly Thege u. 15-17., 1121 Budapest, Hungary, 
              \email{benko@konkoly.hu}
         \and
             Department of Astrophysical Sciences, Princeton University 
             Peyton Hall, 4 Ivy Lane, Princeton, NJ 08544, USA
             \and
             {ELTE E\"otv\"os Lor\'and University, Institute of Physics and Astronomy, 1117, P\'azm\'any P\'eter s\'et\'any 1/A, Budapest, Hungary}
             }

   \date{Received October 30, 2024; accepted March 22, 2025}
 
  \abstract
   {It has long been known that if the durations of the consecutive cycles of a pulsating star vary randomly, the O$-$C diagram could show quasi-periodic/irregular variations, even though the actual average period is constant. It is hypothesised that the period variation observed in many RR Lyrae stars, which are much faster and stronger than may be explained by an evolutionary origin, may in fact be caused by this cycle-to-cycle (C2C) variation effect. So far, quantitative studies have not really been performed, and space data have not been used at all.
   }
   {Our primary goal was to quantitatively analyse the O$-$C diagrams of RR Lyrae stars obtained from space photometry and explained by quasi-periodic or irregular periodic variations to see if they can be explained by random fluctuations in pulsation cycle length without assuming real period variations.  
   }
   {We fitted statistical models to the O$-$C diagrams and tested their validity and fit. The necessary analysis of the light curves was performed using standard Fourier methods. 
   }
   {We found that the vast majority of the O-C curves can be satisfactorily explained by assuming timing noise and the C2C variation without a real mean period variation. We have shown that the strength of the C2C variation is strongly dependent on the pulsation period and metallicity. These correlations suggests that turbulent convection may be behind the C2C variation. The additional frequencies of some RR Lyrae stars and their variation over time play only a marginal role in O$-$Cs. We have found new arguments that the phase jump phenomenon in RRc stars is in fact a continuous change, moreover, it could also be caused by the C2C variation.
   }
   {}


   \maketitle

\section{Introduction}

The period changes of variable stars have been studied for a long time. 
This is also true for classical radially pulsating variable stars such as Cepheids or RR Lyrae stars.
These studies may be aimed at finding period changes due to stellar evolution, detecting possible 
companions, etc. These variations are long time-scales and/or are strictly periodic. 
However, pulsating variables are not precise clocks. 
Their consecutive pulsation cycles could slightly be different from each other.
That this might be the case has been suspected for a very long time, but for most variable types 
it could only be proven by the latest, precise space photometric measurements.

The random period changes were investigated first by \citet{Eddington1929}. 
For long-period Mira variables, it was soon shown that successive pulsation cycle durations actually fluctuate randomly by 1-2\% \citep{Plakidis1932, Plakidis1933}.
For shorter period variables, however, we had to wait for space photometry to find similar variations.
\citet{Derekas2012} detected a period jitter on V1154 Cyg the only 
classical Cepheid star in the original \textit{Kepler} mission \citep{Borucki2010}. 
Then, the cycle-to-cycle (C2C) variation were shown for further different sub-types of Cepheids in the 
data of \textit{K2} \citep{K2}, \textit{MOST} \citep{Evans-MOST-2015}, and \textit{TESS} \citep{Ricker2015} missions \citep{Plachy2017, Plachy2021}.
RR Lyrae stars have shorter periods than Cepheids and therefore only high-cadence measurements 
are suitable for the direct detection of the C2C variation effect, even in space photometry,
as this is the guarantee for a proper coverage of the light curve.
Using 32-second (over-sampled) data of \textit{CoRoT} \citep{Baglin2006} and 1-minute short cadence (SC) 
observations of \textit{Kepler} space telescopes, C2C light curve variations were definitively found 
for non-Blazhko RRab stars by \citet{Benko2016,Benko2019}. 

Traditionally, O$-$C diagrams have been used to detect period changes.
The basic principle is simple:  let $C_i$ be
the time of the $i$th well-defined phase of a light curve
(e.g. maximum, midpoint of the ascending branch, minimum),  
calculated by taking $i$ cycles from a starting time $t_0$ with a constant period $P$.
The actual observed time of the same phase is denoted by $O_i$.
Plotting the difference $(O-C)_i$ as a function of time gives the O$-$C diagram.
For a more detailed overview of the method, see \citet{Sterken2005}. 
Using $O-C$ diagram, we are actually exploiting the fact that a period change always causes 
a change in the observed phase.
If the period $P$ varies in time $t$, it is easy to see from the definition of O$-$C that
\begin{equation}
    (O-C)(t)=\frac{1}{\langle P \rangle }\int_{t_0}^{t} P(t') dt' - t,
\end{equation}
where $\langle P \rangle$ means the average period.
This formula illustrates the cumulative nature of O-C diagrams which helps to detect small 
changes by accumulating them over many pulsation cycles.

It is true that period changes cause changes in the observed phase and hence O$-$C changes, but the O$-$C diagram could also show systematic changes when the average period is constant.
This is a long-known phenomenon. \citet{Sterne1934} was perhaps the first to point out that a random period fluctuation could cause O$-$Cs changes that could be described by a random walk process. 
\citet{Balazs-Detre1965} have already used this phenomenon to explain the strong, irregular O$-$C variations observed in many RR Lyrae stars, which cannot be explained by evolutionary origin. Subsequently, a series of papers have discussed in detail the various mathematical and practical aspects of the problem \citep{Lombard1993, Koen1993, Koen1996, Koen2006}.

Recently, O$-$C diagrams of RR Lyrae stars from space photometric data series have been published, showing quasi-periodic and irregular variations in addition to periodic ones \citep{Benko2019, Benko2023}. In this paper, we address the question to what extent these phenomena can be related to C2C variations of the light curves, hoping to get closer to the origin of these variations.

\section{The method}

\citet{Koen2006} has shown that a quantitative analysis of the O$-$C diagrams can discriminate between
the individual effects: the noise due to the phase determination error, the C2C variation of the light curve, and the actual period variation. The steps of the analysis are described in detail in \citet{Koen2006}, only the necessary information is briefly summarised here.
Koen's statistical model for the instantaneous period $P_i$ of the $i$th pulsation cycle is:
\begin{equation}
    P_i=\langle P \rangle +\sum_{k=1}^{i}\xi_k+\eta_i,
\end{equation}
where $\langle P \rangle$ is the constant average period; the sum represents a general model for
the smooth period variation, where $\xi_i$ are zero mean uncorrelated random variables with variance $\sigma^2_{\xi}$; 
while the C2C variation is modelled by a zero mean uncorrelated random variable $\eta_i$ with
variance $\sigma^2_\eta$. Furthermore, we assume that the $e_i$ error in the timing of the phases is random, with a constant variance $\sigma^2_e$.
\begin{equation}
    t_i=T_i+e_i,  \ \ \ i=0,1,2\dots,k+1,
\end{equation}
where $T_i$ means the error-free timing value.  
The statistical model therefore depends on three random variables 
($\xi$, $\eta$, and $e$) and three associated parameters ($\sigma_\xi$, $\sigma_\eta$, and $\sigma_e$).
In this framework the O$-$C curves can be described in a fairly general way as    
\begin{multline}\label{eq:O-Ci}
    (O-C)_i=e_i + \left(\frac{N_i}{N}-1\right)e_0 - \frac{N_i}{N}e_{k+1} + \\
    \sum_{j=1}^{N_i}\left(\eta_j+\sum_{l=1}^j\xi_l\right) -
    \frac{N_i}{N}\sum_{j=1}^{N_i}\left(\eta_j+\sum_{l=1}^j\xi_l\right),
\end{multline}
where $N_i$ and $N$ means the $i$th and the total ($N=N_{k+1}$) cumulative cycle numbers, respectively.
Using relation (\ref{eq:O-Ci}) and applying different $\sigma$ parameters, curves very similar to the observed 
irregular O-C diagrams can be generated (see figs. in \citealt{Koen2005, Koen2006}). 

When modelling our observed O$-$C diagrams, we consider four scenarios for the possible phase changes.
Since the determination of the data points is always subject to error 
the phase (or timing) noise is unavoidable: $\sigma_e > 0$.
For the first model (M1), we assume that only this timing noise is present, 
nothing else ($\sigma_\eta=\sigma_\xi=0$). In the second case
(M2), we suppose that although the mean period does not change ($\sigma_\xi = 0$), there is a random C2C variation in the light curve causing variation of the pulsation cycle duration ($\sigma_\eta > 0$). In the next model 
(M3) the average period varies  ($\sigma_\xi > 0$) but there is no C2C 
variation ($\sigma_\eta =0$). 
And finally (M4), we allow period variation and C2C fluctuation simultaneously
($\sigma_\eta > 0$ and $\sigma_\xi > 0$).

For each model, we need to maximise the Gaussian log-likelihood function:
\begin{equation}
    {\cal L}_j(\sigma_e, \sigma_\eta, \sigma_\xi)= 
    - \frac{1}{2}\left( k \log\pi + \log\vert \mathbf{\Sigma} \vert + {\mathbf Z}^{\mathsf T}{\mathbf \Sigma}^{-1}{\mathbf Z} \right ), 
    \ \ j=1,2,3,4
\end{equation}
where $\mathbf{\Sigma}=cov( {\mathbf Z}, {\mathbf Z})$ is the covariance matrix of ${\mathbf Z}$ and its elements
are defined as $Z_i=(O-C)_i$. Then we have to
to calculate the Akaike ($AIC$) or Bayesian ($BIC$) information criteria
\begin{align}
    AIC_j & =  - 2{\cal L}_j + 2p + \frac{2p(p+1)}{k-p-1},  \\
    BIC_j & =  - 2{\cal L}_j + p\log k,
\end{align}
where $p$ denotes the number of the free parameters of a certain model ($p=1,2$, or 3).
And the probability of the model is:
\begin{equation}\label{eq:prob}
    p^C_j=e^{-\frac{C_j-C_{\mathrm{min}}}{2}}
    \left( {\sum_{l=1}^4 e^{-\frac{C_l-C_{\mathrm{min}}}{2}}}\right)^{-1},  \ \ \ 
    \mathrm{where} \ \ C=AIC \ \ \mathrm{or} \ \ BIC.
\end{equation}
This procedure leads to the best-fitting (most likely) model. 
When fitting models, \citet{Koen2006} defines 
the pseudo-residual as
\begin{equation}\label{eq:u}
    \mathbf{u}=\mathbf{L}^{-1}\mathbf{Z}, \ \ \ \mathrm{where} \ \ \mathbf{\Sigma}=\mathbf{L L}^{\mathsf T}
\end{equation}
which plays a role similar to that of, for example, residuals in Fourier fits of light curves:
for optimal fitting, the pseudo-residual is a zero-mean, uncorrelated, random variable. 

We have developed {\sc python} code that implements the method described above. 
We have intensively relied on the {\sc NumPy} and {\sc SciPy} extensions of {\sc python}
\citep{NumPy, SciPy}.
The input of the program is the O$-$C diagram, which is formed by a linear transformation if necessary to 
satisfy $(O-C)_0=(O-C)_{k+1}=0$.
The program tests the four statistical models 
maximising the log-likelihood functions with the unknown $\sigma$ parameters.
We tried several bound-constrained maximization algorithms and the Nelder-Mead method of {\sc SciPy} \citep{Nelder-Mead} was
found to be suitable for our problem. 
We calculated the statistical probability (\ref{eq:prob}) of each model using 
both the Akaike and Bayes information criteria.
The program also examines the fit of each model and provides an estimate of the normality of 
the pseudo-residuals using the {\sc scipy.stats.shapiro} routine \citep{Shapiro1965, Royston1995}. 

\section{Statistical models for RRab stars}

\begin{table*}
\caption{\textit{Kepler} non-Blazhko RRab stars}            
\label{table:Kepler_RRab}     
\centering                          
\begin{tabular}{llcccllll}        
\hline\hline                
\noalign{\smallskip}
Name &  $P_0$  &  $\sigma_e$& $\sigma_\eta$ & $p^{AIC}$ & $p^{BIC}$ & Remarks \\ 
     &  (d)  &   ($\times 10^{-5}$~d) & ($\times 10^{-5}$~d) &  &  & \\
\noalign{\smallskip}
\hline
\noalign{\smallskip}
V839 Cyg &       0.4337742 & 3.06 & 0.60 & 0.74 &  0.94 & \\
V368 Lyr &       0.4564859 & 4.10 & 1.04 & 0.73& 0.96 & \\
V715 Cyg &       0.4707059 & 5.32 & 0.98 & 0.85 & 0.50 & M4\\
KIC 6100702 &     0.4881452 & 6.93 & 0.63& 0.73& 0.95& ($f_1$) \\
V349 Lyr &        0.5070742 & 7.15 & 1.61 & 0.66& 0.94& \\
V782 Cyg &       0.5236375 & 7.81 & 1.14& 0.73& 0.96& \\
FN Lyr$^*$ & 0.5273986 & 3.32 & 1.47& 0.73& 0.96&  ($f_1$)\\
KIC 9658012 &     0.533195 & 9.29& 1.93& 0.73& 0.93& $f_2$ \\
V784 Cyg &       0.5340947 &7.65 & 1.06 & 0.73& 0.96& \\
V2470 Cyg &       0.5485897 & 6.51& 0.94& 0.73& 0.96& $f_1$, aRRd? \\
KIC 9717032 &     0.556908 & 6.03& 1.65& 0.73& 0.33& M4\\
V1107 Cyg &       0.5657795 & 6.34& 1.48& 0.73& 0.96& \\
V894 Cyg &        0.5713865 & 6.95& 2.63& 0.74& 0.94& $f_2$\\
V1510 Cyg$^*$  &      0.5811426 & 3.98& 3.37& 0.69& 0.95& $f_2$ \\
NQ Lyr &  0.5877889 & 5.12& 2.05& 0.73& 0.96& $f_1$, ($f_2$), aRRd?\\
NR Lyr$^*$ &  0.6820268 & 7.41& 2.36& 0.73& 0.95& ($f_1$) \\
KIC 7030715 &    0.6836125 & 7.08& 2.22& 0.73& 0.95& \\
AW Dra & 0.6872186 & 4.92& 4.42& 0.63& 0.92& \\
\hline                                   
\end{tabular}
\tablefoot{
The columns of the table contain, in order, the names and pulsation periods $P_0$ of the analysed stars; 
the fitted $\sigma_e$ and $\sigma_\eta$ parameters of the best statistical model and their probabilities 
$p^{AIC}$ and $p^{BIC}$ using the Akaike and Bayesian information criteria, respectively.  
From the O$-$C data of the stars marked with an asterisk, one or two significant short-period signals 
were pre-whitened. $f_1$ and $f_2$ are the additional frequencies published by \citet{Benko2019}, 
identified with the frequencies of the first and second radial overtones.
}
\end{table*}

A significant fraction of RR Lyrae stars show a characteristic simultaneous amplitude and frequency variation with a period (or periods) much longer than the pulsation period. For a detailed discussion of this so-called Blazhko effect, see \citet{Smolec_Blazhko} or \citet{Kovacs_Blazhko2016}. The effect usually dominates the O$-$C diagram of these stars, showing (multi)periodic signal shapes \citep{Benko2014}. Throughout this paper we will focus on non-Blazhko stars, as we are now mainly interested in non-periodic phenomena. 

\subsection{O--C curves of RRab stars}

When \citet{Benko2019} detected the C2C light curve variation of non-Blazhko \textit{Kepler} RRab stars directly, 
the amplitude and light curve shape variation were found to sufficient for detecting such C2C variation.
It is important to point out that variation in the shape of the light curve does not necessarily imply an O$-$C variation. For this to happen, the light curve must change so that the length of the pulsation cycle changes, which can then be detected on the O$-$C diagram.  However, experience shows that the C2C light curve changes of pulsating variables always occur in this way.
The O$-$C diagrams calculated from the \textit{Kepler} long cadence (LC, $\sim 29.5$~min) data and their slightly transformed forms have also been published (see fig.~13 and 14 in \citealt{Benko2019}) and their Fourier analyses have been performed.
Since most of the found frequencies appeared in more than one star, it was thought that they 
might be instrumental frequencies.
However, \citet{Li2023} presented several valuable arguments against these identifications: (i) from a statistical analysis of these frequencies (see their fig.~7), it is concluded that the identification of all such frequencies with instrumental frequencies is unlikely, and (ii) the slopes of O$-$C curves containing similar (considered identical within error) frequencies are also very different, which would not be the case for an instrumental effect. 

We excluded V346\,Lyr from the sample after it turned out to be a Blazhko star, and we tested the remaining 18 stars with the statistical model described above. We used the original O$-$C curves, that is, in contrast to the diagrams presented in \citet{Benko2019}, we have not transformed out any longer-term changes. On the contrary, the four significant high-frequency ($f>0.01$~d$^{-1}$) signals published for NR\,Lyr, FN\,Lyr and V1510\,Cyg are pre-whitened from the data, since these signals are definitely related to some periodic phenomenon and as such are not the subject of the present study.

The results of the analysis are summarised in Table~\ref{table:Kepler_RRab} 
and Fig.~\ref{fig:res_plot_rrab}.
The name and main pulsation period of the star are given in the first two columns of Table~\ref{table:Kepler_RRab}.
The next two columns contain the $\sigma_e$ and $\sigma_\eta$ parameters of the optimal model.
The optimal model is M2 in 16 out of 18 cases (89\%), that is, the O$-$C curve can be described by C2C variation and phase noise.
In two cases, (for V175\,Cyg and KIC\,9717032) the combined M4 model proved to be the best. In both cases, there is a strong linear trend in the O$-$C curves. However, a very similar change can be seen in the O$-$C curves of V368\,Lyr, V1510\,Cyg and AW\,Dra, but for these stars the M2 model is the most likely. This underlines the need for quantitative analysis of O$-$C curves beyond simple visual inspection.

Note that the probabilities $p^C_j$ obtained for the M2 model are between 0.63 and 0.85, while for the same stars the probability of the M4 model is between 0.15 and 0.37. When calculating the $p^C_j$ probabilities, the use of Akaike and Bayesian information criteria suggests a different model in only one case: for KIC\,9717032, Akaike suggests the model M4 ($p^{AIC}_4$=0.85), while Bayes suggests the model M2 ($p^{BIC}_2$=0.66). In this case, following \citet{Koen1996}, we accepted the model obtained from the Akaike formula. The last column shows the additional frequencies that appear permanently, and in parentheses temporarily on the stars. 

In addition to Table~\ref{table:Kepler_RRab}, Fig~\ref{fig:res_plot_rrab} illustrate the analysis performed. The first columns of the figures show the input data, the untransformed O$-$C curves. The panels in the second column show the values of $u_i$, the pseudo-residuals defined by formula (\ref{eq:u}) for the best model. If a model is optimal, the pseudo-residuals are zero-mean 
uncorrelated random variables. A possible verification of this is shown in the panels of the third column: here we give the normalised distribution of the pseudo-residuals compared to the probability density function of the standard normal distribution with $\sigma=1$ variance. The p-value of the Shapiro-Wilk test is also indicated in each panel. The test is usually used so that if the $p$-value is greater than 0.05 it suggests that the data does not significantly deviate from normality. In our case $0.09< p < 0.96$, it means that all the pseudo-residuals are a good approximation of a random variable with normal distribution. 
In principle, it is possible that the pseudo-residual associated with the optimal model is not white noise. In this case, further tests would be needed to accept or reject the model, but fortunately, such a case did not occur in our samples.

\subsection{C2C variation and the physical parameters of RRab stars}

\begin{figure}
\includegraphics[width=0.45\textwidth]{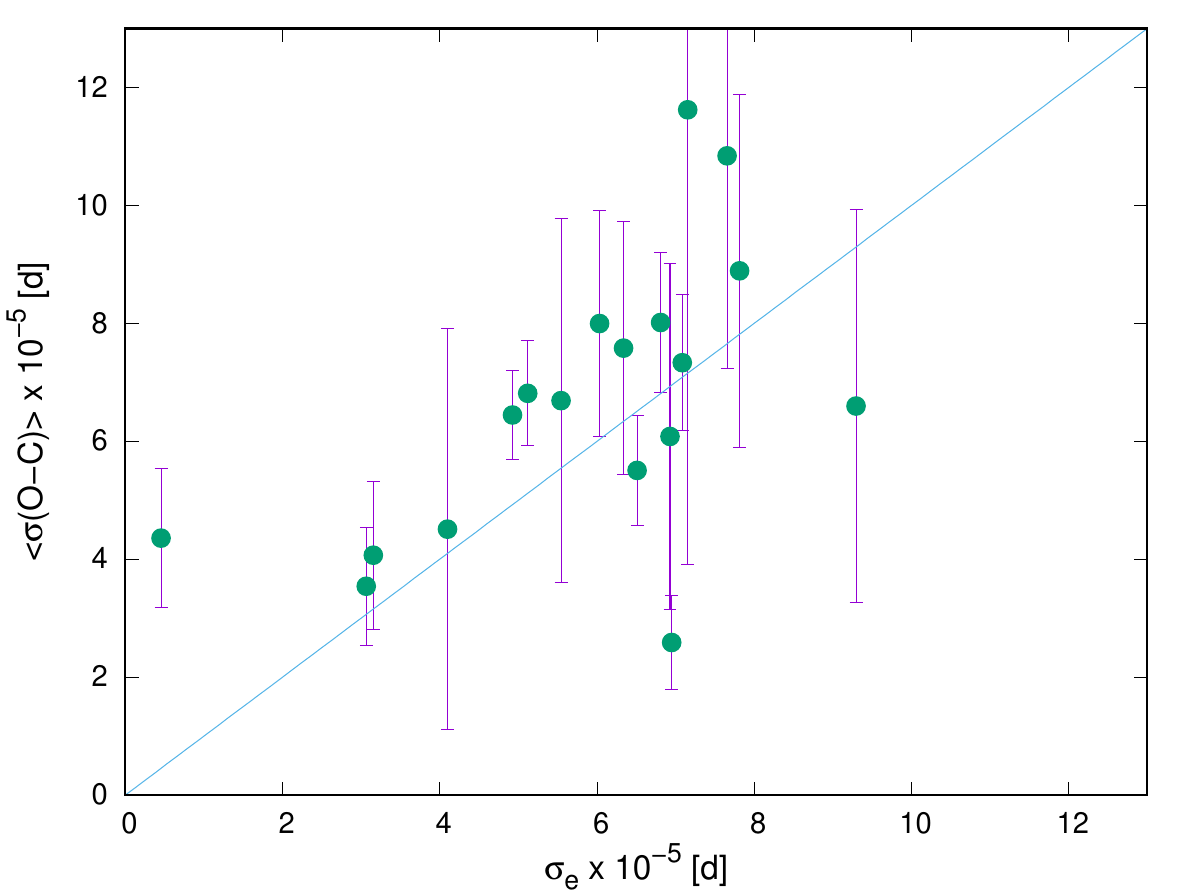}
\caption{Estimations of the phase noise $\sigma_e$ from the 
statistical fit (horizontal scale) with respect to the averaged 
O$-$C determination error (vertical scale) for each RRab star in the sample (dots).
The solid blue line shows a linear with slope one for comparison.
}
\label{fig:sige_rrab}
\end{figure}

\begin{figure}
\includegraphics[width=0.48\textwidth]{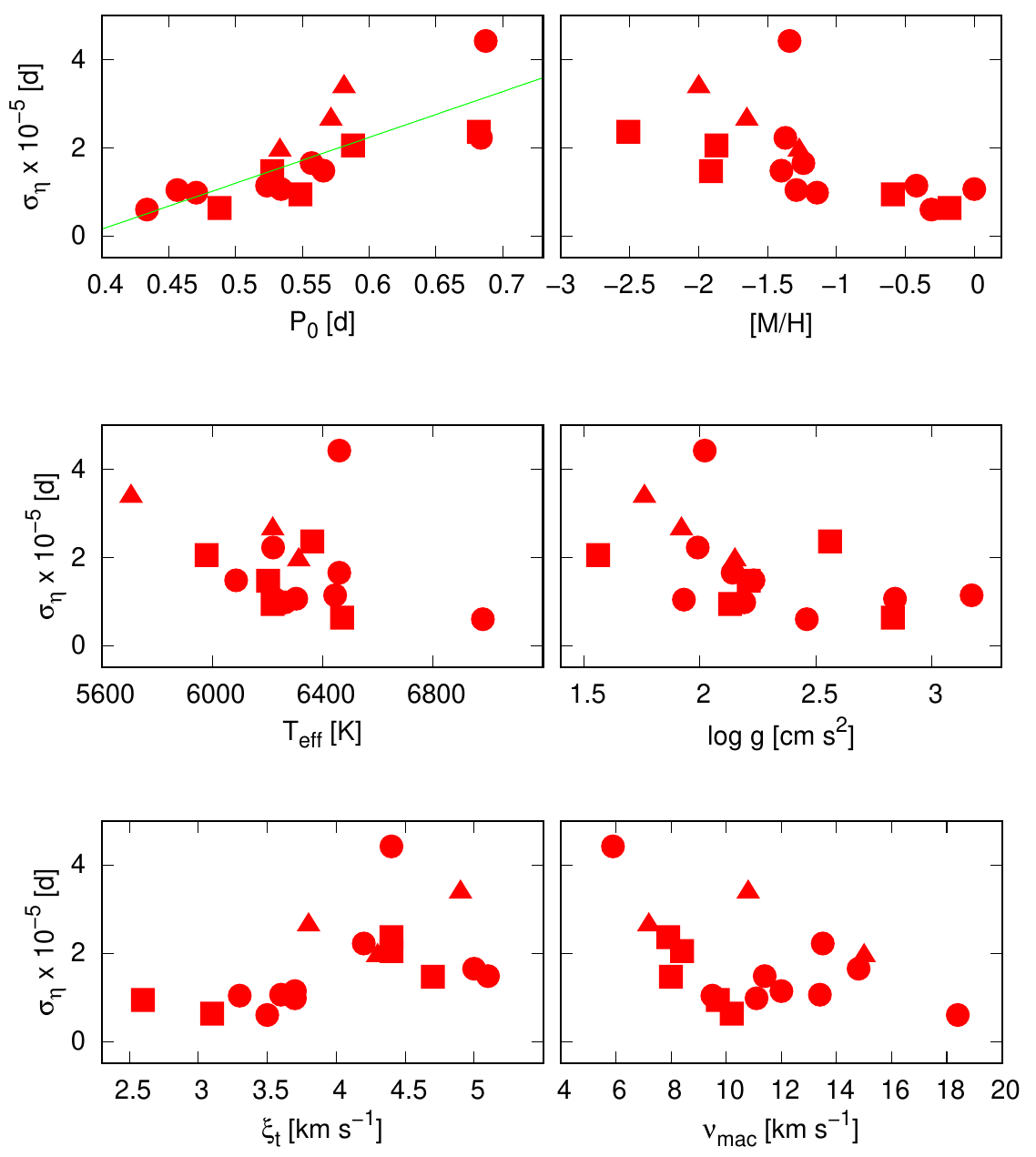}
\caption{The strength of C2C variation as a function of the main period and some physical parameters for \textit{Kepler} RRab stars. 
The stars with no additional frequencies, $f_1$ and $f_2$ frequencies are plotted as filled circles, 
rectangles and triangles, respectively. The green line shows the significant correlation between period and $\sigma_\eta$.
}
\label{fig:fizparam_rrab}
\end{figure}
On the one hand,
our statistical model gives the value of the phase noise $\sigma_e$ for each O$-$C diagram. On the other hand, when the O$-$C values were calculated from the light curves, the error of the individual O$-$C points was also estimated. It may be worth comparing these two error estimates.
We calculate the average and the standard deviation of the individual O$-$C point errors for each star and denote this by $\langle \sigma(O-C)\rangle $ and $\mu_{O-C}$, respectively.
On the horizontal axis of Fig.~\ref{fig:sige_rrab}, we show the $\sigma_e$ values for each RRab star, while on the vertical axis we have shown the above defined $\langle \sigma(O-C)\rangle$ and their standard deviation $\mu_{O-C}$ with error bars. As can be seen, the two error estimates give similar values. Within 3$\mu_{O-C}$ they are identical, and for the majority the difference is within 1$\mu_{O-c}$. 

To unravel the origin of the C2C variation, we have plotted the dependence of the $\sigma_\eta$ parameter characterising the strength of the effect on some important physical parameters in Fig~\ref{fig:fizparam_rrab}. In the panels of Fig.~\ref{fig:fizparam_rrab}, we show the dependences on the main period $P_0$, metallicity [M/H], effective temperature $T_{\mathrm{eff}}$, surface gravitational acceleration $\log g$, micro-turbulence $\xi_t$ and macro-turbulence $\nu_{\mathrm{mac}}$.
Except for the period, the other parameter values were taken from Table 9 of \citet{Nemec2013}, 
determined from high-resolution spectra obtained with CFHT and Keck telescopes.
The stars are marked with different symbols according to their additional frequency contents.
The presence of low-amplitude additional frequencies does not exclude stars from the analysis because, as shown by \citealt{Benko2019}, they do not significantly affect C2C detections.
Filled circles indicate stars not showing additional frequencies, 
while stars containing $f_1$ and $f_2$ radial first and second overtone frequencies represented by rectangles and triangles, respectively. The sizes of the symbols indicate the approximate errors.

Some panels in Fig.~\ref{fig:fizparam_rrab} suggest  relations. 
Since the parameters given here are not independent, a standard multiple linear regression was used to find true significant relationships. In practice, we used the statsmodels {\sc python} package \citep{statsmodels}.
First, all six parameters ($P_0$, [M/H], $T_{\mathrm{eff}}$, $\log g$, $\xi_t$, $\nu_{\mathrm {mac}}$) were fitted simultaneously.  
The obtained $p$-value of the F-statistic ($p_F=0.012$) indicates that at least one of the fitted parameters 
has a significant effect on the dependent variable $\sigma_\eta$. 
Looking at the results for each parameter, we see that there is indeed a positive correlation with the period. The probability of the t-statistic is $p_t=0.021$. For other variables we do not see a significant correlation (all other $p_t > 0.05$).
The obtained  linear relationship for the period is
\begin{equation}\label{eq:corr_rrab}
\sigma_\eta=-4.00(\pm 1.24) + 10.40(\pm 2.22)P_0
\end{equation}
and shown by the green line in the top left panel of Fig.~\ref{fig:fizparam_rrab}:
the amplitude of the C2C variation is higher for stars with longer periods. 
Both $\sigma_\eta$ and period $P_0$ are in day units.

As mentioned in the introduction, it has long been known that the 
C2C variation of the longest period pulsating variables (e.g. Miras) is the strongest relative to the pulsation cycle compared to other pulsators (e.g. Cepheids). 
A linear period dependence within a variable type was published also for the case of Mira stars \citep{Koen_Lombard2004}.
\citet{Balazs-Detre1965} suggested that for RRab stars, the effect is strongest at intermediate periods ($0.11$\% between 0.52 and 0.58~d) and smaller ($0.09$ and $0.05$\%) at shorter and longer periods.
These percentages represent $\sim 3.3$ -- $6\times 10^{-4}$~d, which is about an order of magnitude larger than the ones we obtained, but their statistical model was simpler than the one used here, they did not account for phase noise, but attributed the entire random variation to the C2C effect. 

Unfortunately, no significant correlations with the real physical parameters were obtained. One reason for this is probably the small number of elements in our sample. This is shown that if we omit AW Dra data from the fit (that appear to be outliers), we do not get significant correlations with any of the variables, but $P_0$, $T_{\mathrm{eff}}$ and [M/H] give similar probabilities ($p_t\sim 0.16-0.19$).

\section{RRc stars}

\subsection{Possible C2C variation in RRc stars}

\begin{figure}
    \centering
    \includegraphics[angle=270, width=0.45\textwidth]{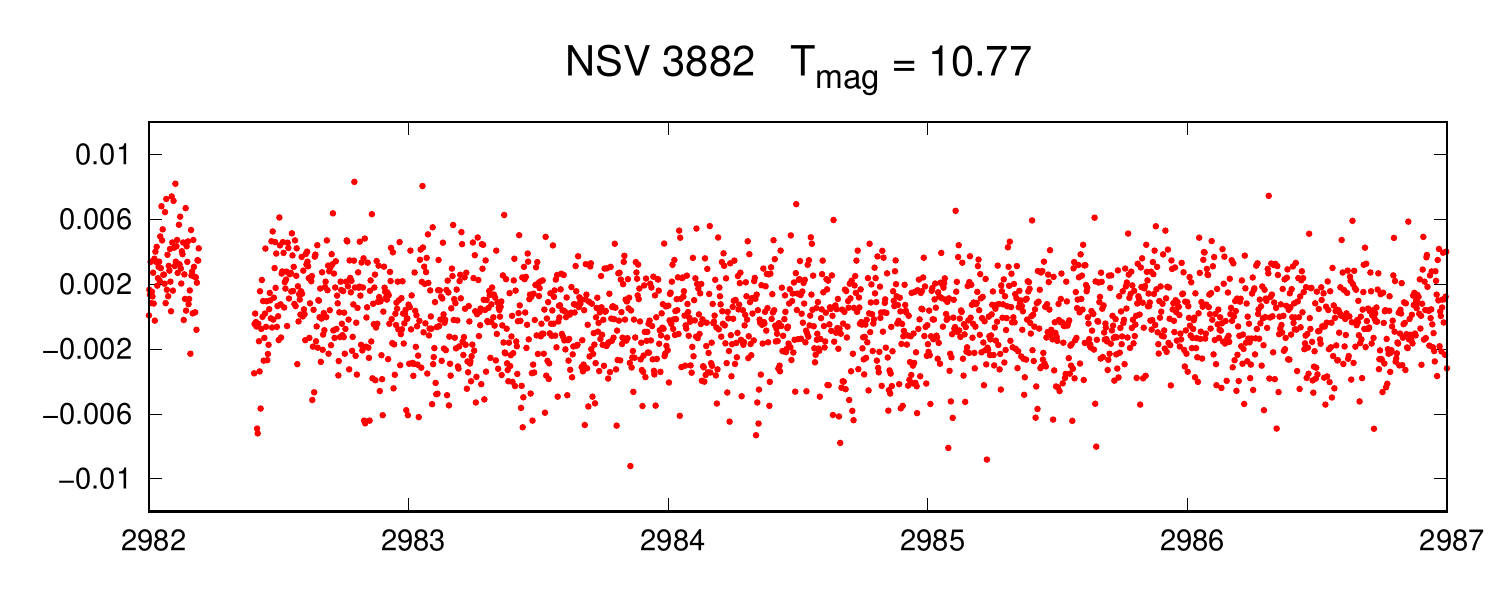}
    \includegraphics[angle=270, width=0.45\textwidth]{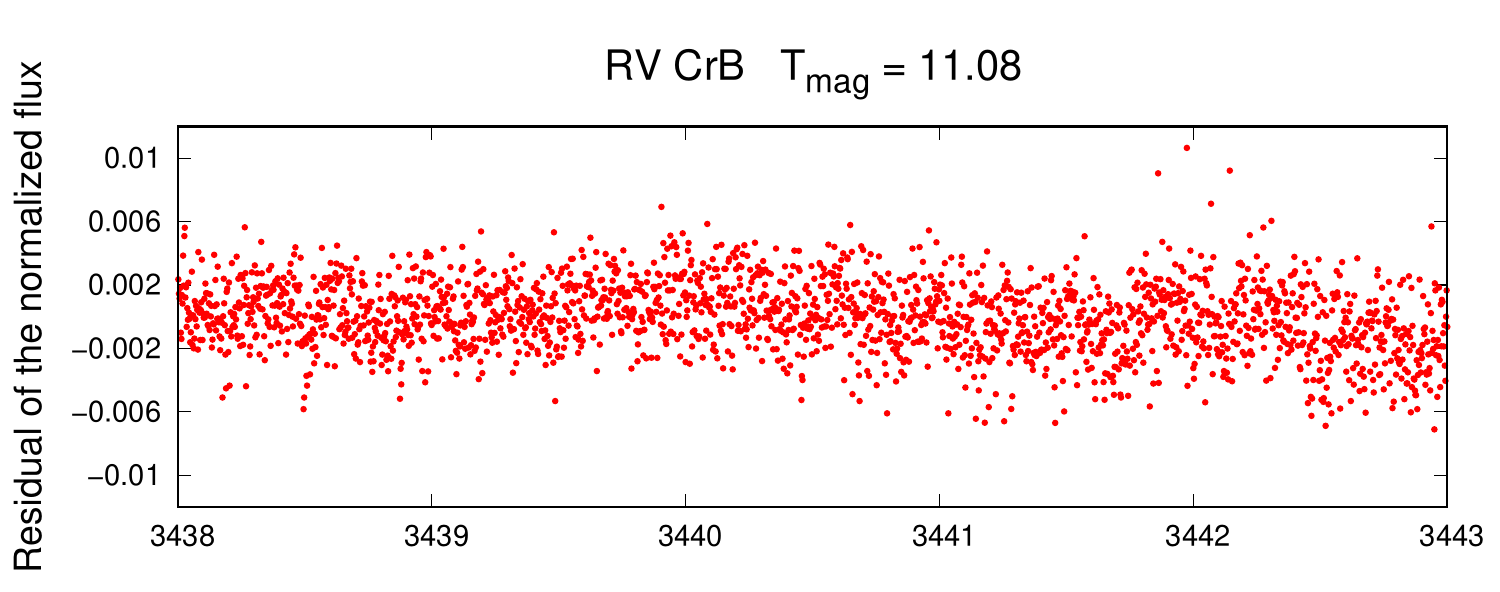}
    \includegraphics[angle=270, width=0.45\textwidth]{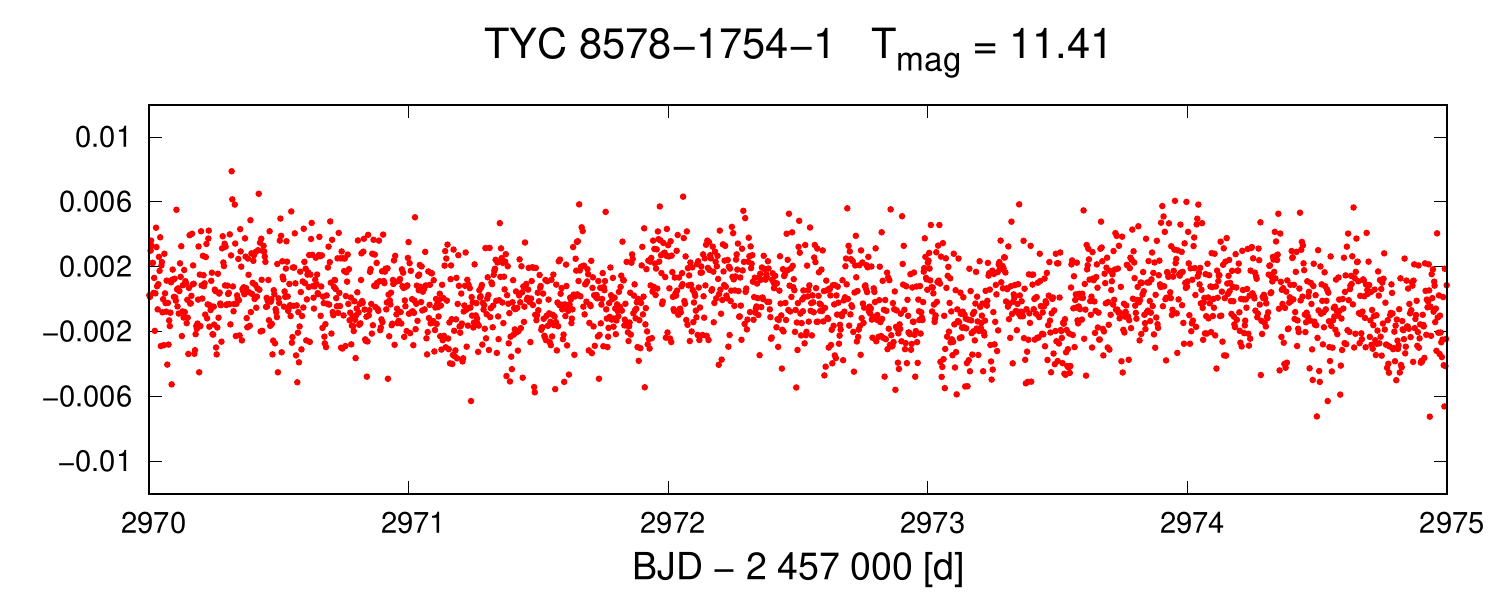}
    \caption{The 5-d length parts of the residual flux curves of the three brightest \textit{TESS} RRc stars
    after the main frequency and its significant harmonics are pre-whitened from the data. The relative flux scale is the same for all three stars. There is no detectable structure indicating a C2C variation.
    }
    \label{fig:RRc_res}
\end{figure}
Direct photometric detection of the C2C variations have not been reported for any RR Lyrae stars that pulsate in the first radial overtone mode (RRc stars). We have now reviewed the available data series and focused on those well-sampled ones in which there is no other variation (caused by additional modes or the Blazhko effect) than the main pulsation, which could interfere with our analysis. 
Although for RRab stars \citet{Benko2019} found that the additional frequencies and C2C variations interact only marginally, this is not known for RRc stars, as the additional frequencies in these stars are of different types and have larger relative amplitudes compared to the main frequency. 
Therefore, to detect C2C directly, it is necessary to first look for `clean' cases, free of additional frequencies.
On the additional frequency content of RR Lyrae stars and their differences by subtypes, see e.g. 
the review articles \citet{Molnar2017} or \citet{Plachy2021}.
  
Checking the available high cadence data from \textit{CoRoT} \citep{Baglin2006} and the original \textit{Kepler} mission \citep{Borucki2010}, we found no RRc star that did not show an additional frequency and/or the Blazhko effect. The \textit{Kepler} extended mission, \textit{K2} \citep{K2}, however,  observed several such RRc stars in 1-minute short cadence (SC) mode. A review of guest observer (GO) data targeting RR Lyrae SC observations found 34 candidates. The candidates were compared with the results of \citet{Netzel2023} and two stars were identified, showing neither the Blazhko effect nor any additional frequencies. These are: EPIC 212640806 ($K_{\mathrm p}=15.889$~mag) and EPIC 249685909 ($K_{\mathrm p}=12.724$~mag). To produce the best possible light curves, the Automated Extended Aperture Photometry method \citep{Bodi2022} was applied to the original CCD frame parts (i.e. `pixel data') of both stars. 

The light curves for the fluxes were analysed with the Period04 software package \citep{Lenz2005}.
After pre-whitening the data of EPIC 212640806  with the main frequency and its significant harmonics,
side peaks appear around the harmonics in the Fourier spectrum. From the residual light curve it is clear that these side peaks are not caused by the Blazhko effect, but by some other effect, such as an instrumental problem.
Otherwise, there is no structure in the residual light curve, only noise. This is consistent with the fact that C2C variations can only be detected up to a certain brightness limit \citep{Benko2019}. The brighter candidate,  EPIC\,249685909,  was found
to have low amplitude but significant frequencies in its residual spectrum after a more detailed analysis. In other words, we have not found any RRc stars in \textit{K2} that are suitable for direct detection of C2C variation.

The initial half-hour exposure time of the full-frame images of the \textit{TESS} space telescope \citep{Ricker2015} gave inadequate phase coverage
for our purposes. Although five non-Blazhko RRc stars were measured by \textit{TESS} at the request of the RR Lyrae working group GO proposal with a sample of 120 seconds, unfortunately, they all contain additional frequencies.
However, after the second extension of the mission (from 1 September 2022), the default exposure time has been reduced to 200 sec. This sampling density may be sufficient. 
From the \textit{TESS} RRc sample of \citet{Benko2023}, we selected stars that did not show any extra signals beyond the main frequency and their harmonics. We found 27 such stars brighter than 13.5 mag, for which the Quick Look Pipeline photometry is available \citep{Huang_1_2020, Huang_2_2020, Kunimoto_1_2021, Kunimoto_2_2022}. These photometric data series were examined to look for signs of C2C variation. 

Taking into account the difference in apertures (10.5 cm for \textit{TESS}, 95 cm for \textit{Kepler}) and exposure times (58.9~sec for \textit{Kepler} SC, 200~sec for \textit{TESS}), we estimated the detection limit of \textit{TESS} to be about 3.5 mag brighter than \textit{Kepler}. For the \textit{Kepler} RRab stars the detection limit of the effect was found to be $K_p\sim 15$~mag \citep{Benko2019}
which means that only the brightest \textit{TESS} RRc stars are candidates for the detection of the C2C effect. The C2C variation are well traced on the residual light curves from which the main frequency and its significant (10-15) harmonics have been pre-whitened as illustrated in figures 1 and 2 in the paper by \citet{Benko2019}. Figure~\ref{fig:RRc_res} shows the 5 day chunks of the residual light curves of the three brightest stars in our sample. Unfortunately, none of them show any definite structure.
In the case of RRab stars, because of the different strengths of the C2C variation in different phases of the light curve, we can see a definite structure in the residuals (see fig.~1 and fig.~2 in \citealt{Benko2019}). 

Thus, it appears that there is currently no photometric data set from which the C2C variation of RRc stars 
can be directly detected. However, indirect analysis of C2C variations using O$-$C diagrams can be done 
in a similar way to RRab stars.

\subsection{O$-C$ diagrams of RRc stars}

\citet{Benko2023} published phase variations on a one-year timescale for 32 RRc stars (see their fig.~14) in the \textit{TESS} Continuous Viewing Zone (CVZ). From these phase variations we constructed the O$-$C diagrams for the 26 non-Blazhko stars. Similar to the RRab stars, the evident short period signals with the frequency of $f > 0.03$~d$^{-1}$ were also removed. From the O$-$C curves of each star, 1-5 (on average 2) such frequencies  were pre-whitened (for the frequency values see Table 6.\ in \citealt{Benko2023}). The resulting curves were used to fit the statistical models described above for RRab stars. The O$-$C diagrams, the pseudo-residuals corresponding to the best-fitting models, and their distribution are shown in Fig.~\ref{fig:res_plot_rrc}. The numerical results are summarised in Table~\ref{table:TessRRc}. 

Remarkably, the value of $\sigma_\eta$ is on average an order of magnitude larger than for RRab stars, that is  $\langle\sigma(\mathrm{RRc})_\eta\rangle=1.20\times10^{-4}$ vs.  $\langle \sigma(\mathrm {RRab})_\eta\rangle=1.76\times10^{-5}$. The study of \citet{Balazs-Detre1965} gave 0.04\% for the estimated C2C period variation of RRc stars, which means  $\sigma_\eta \approx1.0 - 1.6\times10^{-4}$, and thus in agreement with the result obtained here. 

\begin{table*}
\caption{\textit{TESS} RRc stars}            
\label{table:TessRRc}      
\centering                          
\begin{tabular}{lllllllllllll}        
\hline\hline                
 \noalign{\smallskip}
Name &  $P_1$  & $\sigma_e$& $\sigma_\eta$ & $p^{AIC}$ & $p^{BIC}$ & $T_{\mathrm{eff}}$ & $\log g$& [Fe/H] & Remarks \\ 
     &  (d)  &  ($\times 10^{-4}$~d) & ($\times 10^{-4}$~d) &   &  & (K) & (cm s$^2$)& &  \\
      \noalign{\smallskip}
\hline
 \noalign{\smallskip}
$[$SHM2017$]$ J282.06827+62.45832   & 0.249948 & 3.19 & 0.17   & 0.41 & 0.55  & 7140 & 3.92 & $-$0.63 &  $f_{61}$ \\
Gaia DR2 5476969341269561344   & 0.268693 &  1.51  &     &  0.63 & 0.79 & 7210 & 4.04 & $-$1.44 & M3, $f_{60}$ \\
Gaia DR2 1650042086261498880   & 0.280923 & 0.91 & 0.56  &  0.97 & 0.81      & 7180 & 2.97$^*$ & $-$1.06 & M4, $f_{61}$ \\
Gaia DR2 5280964179391935616   & 0.282452 & 1.01  & 0.27 &  0.73 & 0.95      & 6940 & 3.88 & $-1.35$ & $f_{61}$\\
$[$SHM2017$]$ J280.68767+58.39747  & 0.292591 & 2.20 & 0.83 & 0.73 & 0.95    & 7010 & 4.12 & $-1.39$ & $f_{61}$\\ 
2MASS J04274606-6202513 &   0.294156 & 1.16 & 0.38  & 0.73 & 0.94            & 7420 & 3.48 & $-1.71$ & \\
Gaia DR2 5264448926328694272  & 0.309862 & 2.02 & 1.04 & 0.73 & 0.95          & 6890 & 4.04 & $-1.35$ & $f_{61}$, $f_{63}$ \\
Gaia DR2 5486632674089049472  & 0.314167 & 0.68 &     & 0.73 & 0.94          & 6820 & 3.86 & $-2.27$ & M3 \\ 
Gaia DR2 4628067852624828672  & 0.316098 & 1.32  &  0.20 &  0.72 & 0.92      & 6900 & 3.87 & $-1.71$ & $f_{68}$ \\
Gaia DR2 5285349822037246464  & 0.321226 & 0.73 &  0.18 & 0.74 & 0.94       & 7010 & 3.88 & $-2.26$ & \\
HD 270239 &  0.321779 &  2.58 &  1.53 & 0.65 & 0.92                          & 8230 & 3.89 & $-1.50$ &  \\
GSC 04450-00308 &  0.326734 &  0.53 & 0.35  &  0.73 & 0.95                  & 6830 & 3.88 & $-1.83$ & \\
$[$SHM2017$]$ J281.12439+66.86821  & 0.337833 &  2.71 & 1.79 & 0.91 & 0.98    &  6810 & 3.85 &  $-1.53$ & $f_{61}$, $f_{63}$  \\
Gaia DR2 5482545510194122112 &  0.337873 & 2.75 & 0.35 & 0.46 & 0.57          &  7270 & 3.84 & $-1.57$ & $f_{68}$ \\
Gaia DR2 4758809680772463232 &  0.34105 &  3.19 & 1.62 & 0.73 & 0.95          &  6870 & 3.83 & $-1.54$ & $f_{61}$, $f_{63}$\\
CRTS J165930.9+650919 &  0.341773 & 2.42 & 1.56 & 0.73 & 0.95                 &  6790 & 3.21 & $-1.60$ &  $f_{61}$, $f_{63}$ \\
Gaia DR2 5264031833464989568 &  0.342671 & 1.72 & 1.62 & 0.74 & 0.94          &  6760 & 3.71 & $-1.58$ &  $f_{61}$, $f_{63}$\\
Gaia DR2 2251757108828618496 &  0.343299 & 2.52 & 0.88 & 0.74 & 0.94 & 6470 &  & $-1.87$ & $f_{63}$, $f_{61}$\\
Gaia DR2 5280883674523697792 &  0.347276 & 5.11 & 1.64 & 0.74 & 0.94 & 6870 &  & $-1.84$ & $f_{61}$, $f_{63}$\\
Gaia DR2 1439840098963073152 &  0.357564 & 2.06 & 1.26 & 0.74 & 0.95 & 6880 & 3.78 & $-2.76$ & $f_{63}$, $f_{61}$\\
NSVS 3117163 &  0.35856 & 1.97 & 1.17 &  0.72 & 0.94 & 6730 &  & $-1.97$ & $f_{61}$, $f_{63}$\\
$[$SHM2017$]$ J296.89770+71.59565 & 0.359776 & 2.47 & 0.33 & 0.67 & 0.91 & 6740 &  &  $-2.31$ & $f_{60}$\\
Gaia DR2 5285510178935677312 &  0.385039 & 3.27 & 5.11 & 0.89 & 0.61 & 6850 & 3.82 & $-0.38$ & M4, $f_{61}$, $f_{63}$ \\
CRTS J165523.7+573029 &  0.392964 & 2.18 & 2.16 & 0.74 & 0.95 & 6710 & 3.25$^*$ & $-2.49$ & $f_{61}$, $f_{63}$\\
$[$SHM2017$]$ J285.32992+67.89109 &  0.39449 & 1.64 & 1.73 & 0.99 & 0.92 & 6940 & 4.04  & $-2.41$ & M4, $f_{61}$, $f_{63}$ \\
CRTS J165435.7+655131 &  0.406081 & 1.75 & 2.09 & 0.74 & 0.94 & 6630 & 3.97 & $-2.65$ & $f_{61}$, $f_{63}$ \\
\hline                                 
\end{tabular}
\tablefoot{
The columns of the table contain, in order, the name and main pulsation period $P_1$; 
the fitted $\sigma_e$ and $\sigma_\eta$ parameters of the best statistical model and their probabilities 
$p^{AIC}$ and $p^{BIC}$ using the Akaike and Bayesian information criteria, respectively;  
some physical parameters: effective temperature ($T_{\mathrm{eff}}$), surface gravitation acceleration ($\log g$),
metallicity ([Fe/H]). Last column shows the additional frequencies detected and denoted\footnote{
In their notation, the ratio of their frequencies is represented by the main frequency $f_1$: i.e. $f_x:=f_{61}$ if $f_1/f_x \approx 0.61$; $f_x:=f_{63}$ if $f_1/f_x \approx 0.63$; but $f_x:=f_{68}$ if 
$f_x/f_1 \approx 0.68$.} 
by \citet{Benko2023} and the name of the optimal statistical model where it differed from the M2. The two $\log g$ values marked with asterisks are from the LAMOST Survey \citep{Lamost_RR}.
}
\end{table*}

For 21 of the 26 RRc stars, the optimal model, M2 (81\%), that is the O$-$C curve, can be described by the C2C variation and the phase noise. For three cases (Gaia DR2 1650042086261498880, Gaia DR2 5285510178935677312 and  [SHM2017] J285.32992+67.89109) the combined (M4) model proved to be the best, while for two stars (Gaia DR2 5476969341269561344 and Gaia DR2 
5486632674089049472) the combination of phase noise and a real period variation was found to be the most probable.

\subsection{C2C variation and the physical parameters for RRc stars}

For RRab stars, we found correlation between the strength of the C2C variations and the main pulsation period. We attempted to find correlations for the RRc sample. 

Unfortunately, high-resolution spectroscopy is not available for our RRc stars. Therefore, we used the calibration of \citet{Li_FeH2023}, who determined [Fe/H] values from the \textit{Gaia} DR3 $G$-band photometry. Since not all of our stars were included in their table, we used the 
\begin{equation}
    \mathrm{[Fe/H]}=-1.737 - 9.968(P_1-0.3) - 5.041(R_{21}-0.2)
\end{equation}
formula they gave us. Here the $P_1$ is the mean period calculated from the \textit{TESS} light curves and $R_{21}=A(2f_1)/A(f_1)$ is the amplitude ratio of the $G$-band light curves. For the stars in the \citet{Li_FeH2023} table, our calculated and published [Fe/H] values agree within a few hundredths of a percent, well below the 1$\sigma$ error limit ($\sim$0.2-0.3~dex). Used [Fe/H] values are listed in col. 9 in Table~\ref{table:TessRRc}.

\begin{figure}
\includegraphics[width=0.45\textwidth]{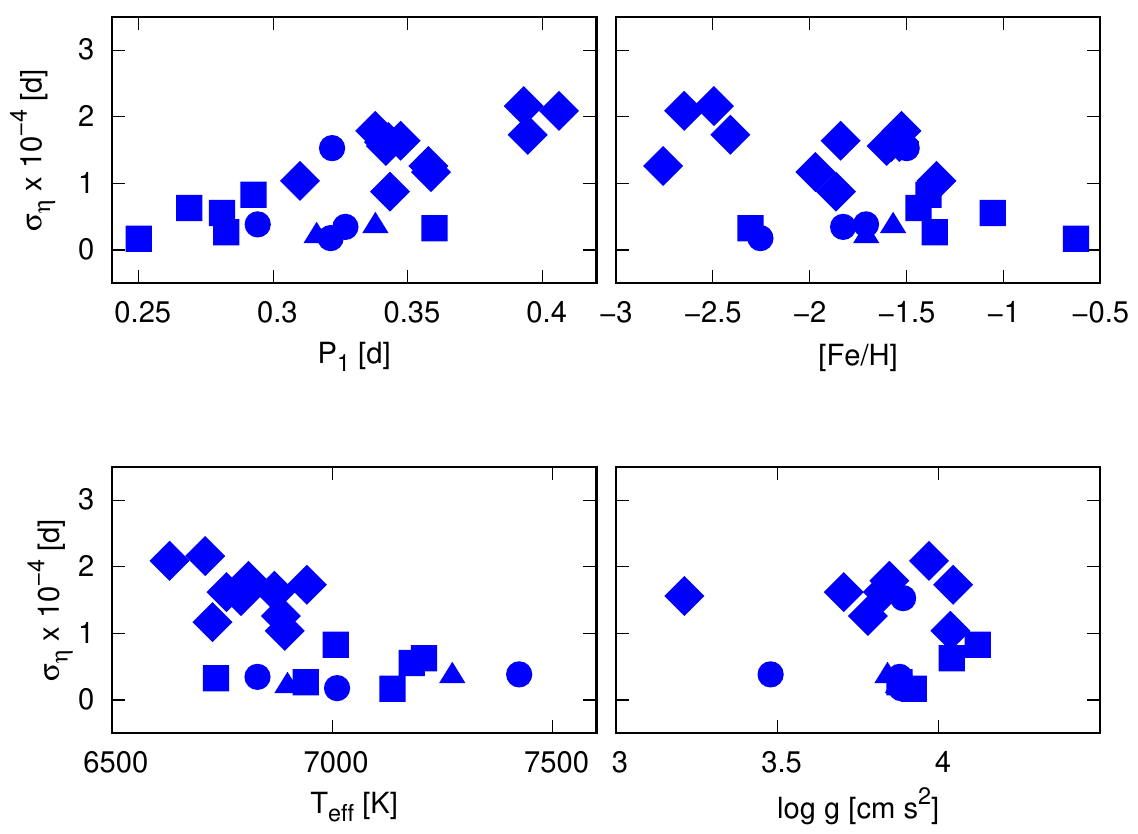}
\caption{
The parameter $\sigma_\eta$ characterizing the C2C variation for the stars of the RRc sample as a function of the main pulsation period $P_1$, metallicity [Fe/H], effective temperature $T_{\mathrm {eff}}$, and $\log g$.
The different symbols show the additional frequency content of the stars (see text for the details).
}
\label{fig:fizparam_rrc}
\end{figure}

The effective temperatures were determined by the algorithm of \citet{Casagrande2021} using \textit{Gaia} DR3 and 2MASS colours and shown in col.~7 of Table~\ref{table:TessRRc}. The values are rounded to ten, the average error is $\sim 150$~K.

In the most recent publication of the LAMOST Survey of RR Lyrae \citep{Lamost_RR}, only two stars from our sample are included, so the $\log g$ values are taken from \textit{Gaia} DR3 \citep{Gaia_DR3} as the most extensive homogeneous source, even so, we do not have data on six of our stars (col. 8 in Table~\ref{table:TessRRc}).
We used the `GSP-Phot' values estimated from the $BP-RP$ colour indices. 

In the panels of Fig.~\ref{fig:fizparam_rrc}, we plotted the C2C variation strength parameter $\sigma_\eta$ as a function of the main period $P_1$, the metallicity [Fe/H], the effective temperature $T_{\mathrm{eff}}$ and the surface gravitational acceleration $\log g$. The physical parameters used are listed in Table~\ref{table:TessRRc}. The stars are marked with different symbols according to their additional frequency content. Filled circles, squares, triangles or diamonds represent stars with no additional frequency, one additional frequency ($f_{60}$ or $f_{61}$), $f_{68}$ frequency and two frequencies (both $f_{61}$ and $f_{63}$), respectively. 

To search for relations, as for the RRab stars, we first performed simultaneous multiple linear regressions with all four parameters: $P_1$, [Fe/H], $T_{\mathrm{eff}}$, $\log g$. Strong ($p_t < 0.001$) significant linear dependence with period and metallicity were obtained according to the following formula:
\begin{equation}\label{eq:corr_rrc}
    \sigma_\eta=-5.67(\pm 0.86)+27.87(\pm 3.04)P_1+1.42(\pm 0.21)\mathrm{[Fe/H]}.
\end{equation}
$\sigma_\eta$ and $P_1$ are measured in days and [Fe/H] in the usual dimensionless quantity, dex.
No correlation was found for effective temperature or $\log g$.
However, we note that stars with two additional frequencies (diamonds in Fig.~\ref{fig:fizparam_rrc}) are grouped at lower temperatures (bottom left panel).
This has already been mentioned by \citet{Benko2023} based on the period distribution of these stars, but has yet not been investigated in detail.
The lack of correlation with $\log g$ may also be due to the known fact that the \textit{Gaia} physical data for horizontal branch stars such as RR Lyrae (cf. Table~\ref{table:TessRRc} data and fig. 14 in \citealt{Molnar_M80}) not really accurate.

\subsection{The `phase jump' phenomenon}

Longer-term ground-based observations of some RRc stars -- observing the stars over several seasons -- have found that while the shape of the light curve appears to be constant, no single period can describe
the light curves. However, assuming that a phase jump occurred in the unobserved time interval, the light curves can be folded perfectly (e.g. \citealt{Wils2007, Wils2008, Odell2016, Berdnikov}). This behaviour is known as phase jump phenomenon.
The long continuous data series of the \textit{Kepler} space telescope suggested that the phenomenon is most likely not the consequence of a sudden jump, but rather the result of a continuous phase change \citep{Moskalik2015, Sodor2017}. For example, in the case of KIC\,4064484, the phase is almost constant for $\sim 100$~d, and then it changes by 0.3~rad on a similar time scale of 100 days.
After that, it stays at this level for about $\sim 200$~d, and then returns to almost the initial value again in 150-200 days (see fig. 9 in \citealt{Moskalik2015}). Such a phase change from the Earth's surface, observed seasonally, could be modelled by at least two abrupt phase jumps. Continuous phase changes similar to KIC\,4064484 can be seen on our several \textit{TESS} O$-$C curves (e.g. CRTS\,J165523.7+573029, NSVS\,3117163, or CRTS\,J165930.9+650919 in Fig.~\ref{fig:res_plot_rrc}).

The above explanation of the phase jump phenomenon was first proposed by \citet{Benko2023}. Here we now take two further steps to understand what causes the observed phenomenon. (i)
The times corresponding to the phase jumps range from a few seconds to tens of minutes, 
that is between $\sim10^{-4}$ and $\sim10^{-2}$~d.
The vertical scales of the O$-$C diagrams in Fig.~\ref{fig:res_plot_rrc} are exactly in this interval. 
In other words, the phenomenon is not only qualitatively possible, but the observed O$-$C values are also quantitatively appropriate. (ii) Since we found that the majority of the O$-$Cs can be explained by the C2C variations of the light curve, we can conclude that the jumps observed in the ground-based data are nothing more than the result of incomplete observations of (continuous) phase changes due to C2C variations. This hypothesis is supported by the fact that the shape of the light curve of phase jump stars (by definition) does not change, at least within the accuracy of ground-based measurements.

\section{Discussion}

As we have seen, 81-89\% of the O$-$C variations in our RR Lyrae sample (16 stars out of 18 RRab and 21 stars out of 26 RRc) can be explained purely by timing noise and C2C variations of the light curves without assuming a real period change.
\begin{figure}
    \centering
    \includegraphics[width=0.48\textwidth]{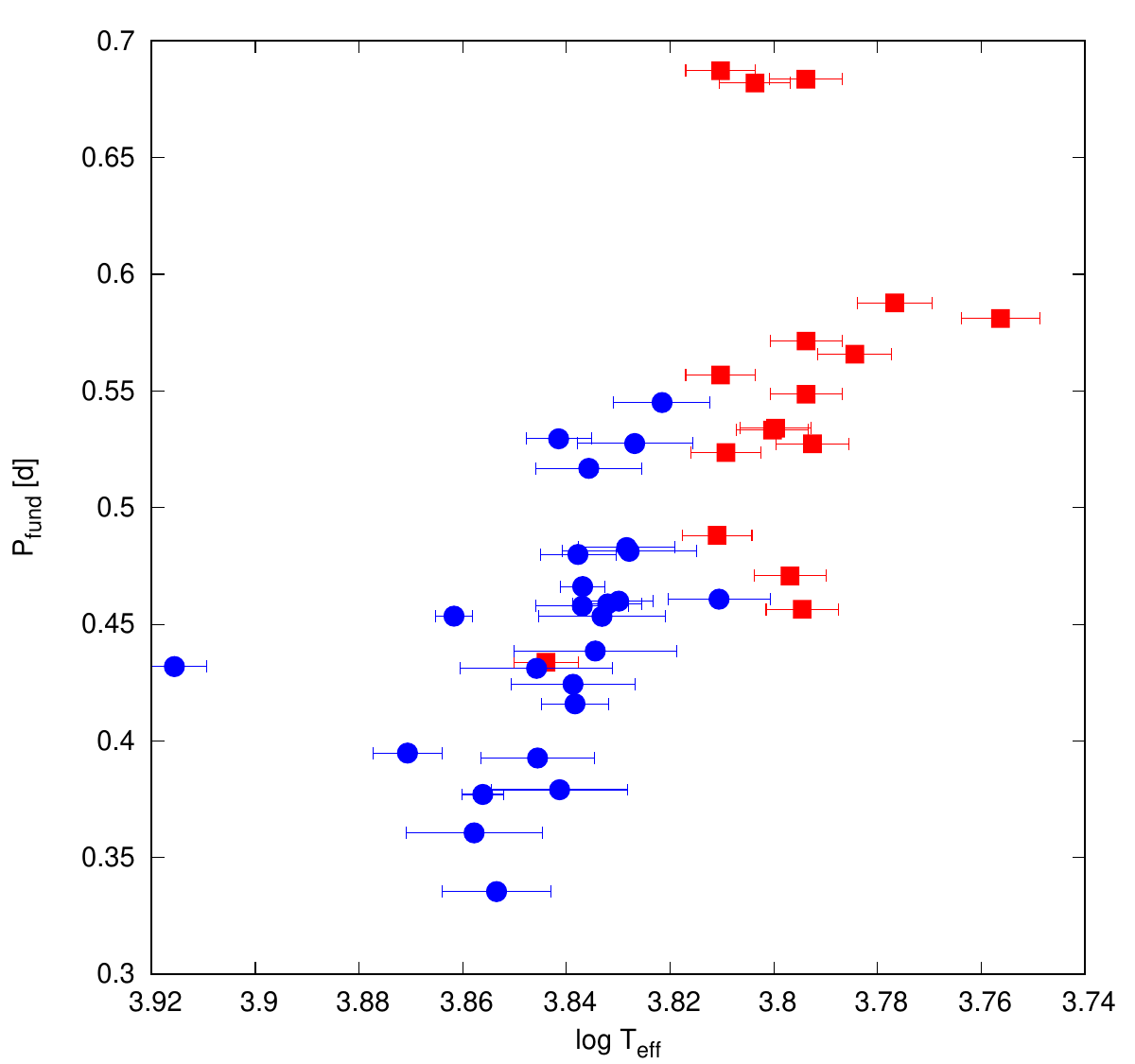}
    \caption{ The relation between the fundamentalised period $P_{\mathrm{fund}}$ and the effective temperature for the stars in our sample. Red rectangles: RRab stars, blue circles: RRc stars. 
    }
    \label{fig:p_teff}
\end{figure}
What is the physical origin of the C2C variation phenomenon? The one-dimensional theoretical pulsation models \citep{Buchler1997, Feuchtinger1999b, Smolec2008} can not reproduce the C2C variations: after reaching the limit cycle these models are strictly repetitive. \citet{Sweigart1979} explained the effect by random mixing events in the semi-convective layer around the convective core. The hypothetical mechanism is as follows: mixing causes a change in the layering of the chemical composition. As a result, the hydrostatic equilibrium of the star changes: the radius of the star, and hence the pulsation period, changes. \citet{Deasy1985} mention several explanations for the random periodic variations in Cepheids: one is the interaction of turbulent convection and pulsation, the other is the mass loss. They have not carried out quantitative studies of either but \citet{Cox1998} has shown via theoretical calculations that small changes in the He gradient at the bottom of the H and He convection zones can indeed cause sufficient period changes.

The correlations between the parameter $\sigma_\eta$, indicating the strength of C2C variation, and the period found in (\ref{eq:corr_rrab}) and (\ref{eq:corr_rrc}) might be explained qualitatively by convection. As shown by \citet{Bellinger2020} the period (and amplitude) of an RR Lyrae star is mostly determined by the effective temperature. Our own analysis showed well that there was a strong multicollinearity in the simultaneous regression of period and effective temperature, showing that these are highly correlated parameters. 
This is illustrated in Fig.~\ref{fig:p_teff}, where we plot the fundamentalised period ($P_{\mathrm{ fund}}=P_0$, or $P_1/0.745$) of the stars in our sample as a function of $\log T_{\mathrm {eff}}$. It can be seen that both the RRab and RRc samples are relatively small individually, but together a strong non-linear relationship is visible. 
Longer periods are typically associated with lower temperatures and vice versa. 
Then, by relating $\sigma_\eta(P)$ functions to temperature, we find that the strength of the C2C variation increases towards lower temperatures. However, we know that the strength of convection also increases with decreasing temperature. 

There is no such obvious explanation for the correlation with metallicity. Metallicity affects the internal structure of the star, and somehow the found correlation might be a consequence of that. 

So the indications suggest that convection
may be behind the C2C variation phenomenon. This is supported by the fact that, despite their shorter pulsation periods and higher effective temperatures, RRc stars show on average an order of magnitude stronger C2C variations. Indeed, it is known that while convection interacting with pulsation is significant only in certain pulsation phases in RRab stars, convection is continuously present in RRc stars 
\citep{Gautschy2019}. The theoretical description of the interaction between pulsation and convection is still a topic of ongoing research \citep{Kovacs2023, Kovacs2024}. The question can only be satisfactorily resolved in the future using multidimensional pulsation codes. Significant initial steps have been made with such codes \citep{Deupree2015, Mundprecht2015, Kupka2017}, but the descriptions are far from complete. Hopefully, these studies will be able to reproduce C2C variations in a natural way. 

To describe the O$-$C curves, we had to assume actual period variation (models M3 or M4) for two RRab stars (11\%) and five RRc stars (19\%) among our studied stars. These period changes are rather slow, with a minimum time scale of half a year.  If these are periodic variations, their period can vary from about a half a year to several years. The measured period variation is between $7.3\times10^{-3}$ and $3.3\times10^{-2}$~d. Assuming periodic variation, these values are lower limits. Several explanations have been proposed for such long-period phase variations. \citet{Li_Qian2014, Li2023} explained the variation of FN Lyr, V894 Cyg and V838 Cyg by low-mass companions around the stars, while \citet{Benko2019} hypothesised it as a instrumental effect. FN Lyr and V894 Cyg are included in our sample, and C2C variations seem to be a satisfactory explanation in both cases. (V838 Cyg shows a Blazhko effect and was therefore omitted by this work.) However, the assumption of binarity or a companion should always be considered for these types of changes. Since in this case we are dealing with a strictly periodic signal, with a time series of sufficient length (covering several complete cycles) such an assumption can be confirmed. If \textit{TESS} does indeed continue its observations for 10-15 years, there is hope of testing the hypothesis for some stars.

\section{Conclusions}

(i) \citet{Benko2019} showed, based on the SC data of \textit{Kepler} non-Blazhko RRab stars, that the light curves of these stars vary slightly from cycle to cycle. In this work, we performed statistical modelling of the O$-$C curves calculated from the LC data of the same stars. On this basis, we show that the vast majority of the O$-$C curves (16 out of 18 stars) can be satisfactorily explained by assuming timing noise and the C2C variation phenomenon without a true mean period change. 

(ii) The statistical fits were not sensitive to the additional frequencies that appear in certain stars. This is consistent with the finding of \citet{Benko2019} that the C2C variation effect dominates over the influence of the additional frequencies. 

(iii) We have shown that the strength of the C2C variation is strongly dependent on the pulsation period: the effect is stronger for longer periods. The strength of the C2C variation in RRc stars depends also on metallicity.
From the nature of the connections, we can suspect that turbulent convection 
may be behind the C2C variation.

(iv) 
The C2C variation of RRc stars could not be detected directly by comparing the successive pulsation cycles of the light curves, as was done for RRab stars, because no data series could be found that met all the necessary criteria. However, the long-term effects of the C2C variation on the O$-$C diagrams could be used to estimate the strength of the C2C variation, as for RRab stars.
This was found to be on average an order of magnitude larger than for RRab stars, but the dependence on pulsation period are similar to the correlations found for RRab stars. Again, for 81\% of the stars, the timing noise and the C2C variation were sufficient to explain the observed O$-$Cs.

(v) Considering the time scale and shape of the O$-$C variation in RRc stars, we suspect that the phase jumps found in ground-based observations are likely due to non-continuous observations of continuous phase variations caused by C2C variation.

\begin{acknowledgements}
This paper includes data collected by the \textit{Kepler} and \textit{TESS} missions. Funding for the missions are provided by the NASA Science Mission Directorate. The research was partially supported by the ‘SeismoLab’ KKP-137523 \'Elvonal grant of the Hungarian Research, Development and Innovation Office (NKFIH) and by the NKFIH excellence grant TKP2021-NKTA-64. Some {\sc python} codes were developed with the help of ChatGPT 3.5. The authors thank the reviewer for her/his helpful suggestions and
J. Bienias for his correction to the language.
\end{acknowledgements}

\bibliographystyle{aa}
\bibliography{oc.bib}

\begin{appendix}
\onecolumn

\begin{figure*}[ht]
\section{Statistical fitting of the O$-$C curves of the non-Blazhko RRab stars of the original \textit{Kepler} mission.}
\subfloat[]{
\includegraphics[width=0.8\textwidth]{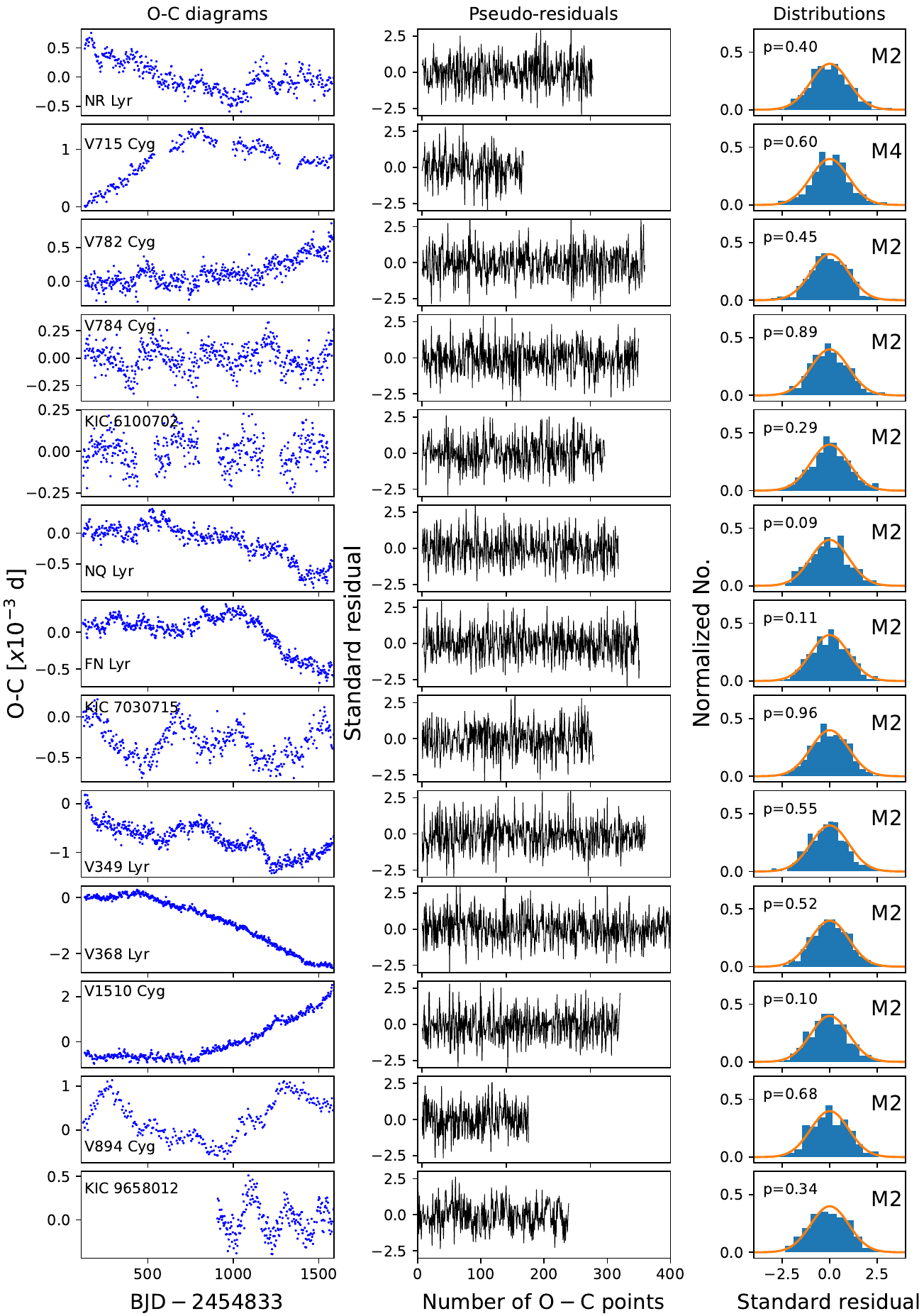}}
\caption{
First column: The `raw' O$-$C diagrams -- without subtracting the quadratic fit performed by \citet{Benko2019}.
Note the very different vertical scales of each panel. 
Second column: The pseudo-residuals (see formula (\ref{eq:u})) associated with the best statistical model.
Third column:  Normalised distribution of pseudo-residual values (blue histograms) compared to the standard normal distribution (orange curves). 
In the upper right corner of the panels, the resulting optimal model (M1-M4) is indicated, in the upper left corner, the p-value of the normality test is given.
}\label{fig:res_plot_rrab}
\end{figure*}
\begin{figure*}
\ContinuedFloat
\subfloat[]{\includegraphics[width=0.8\textwidth]{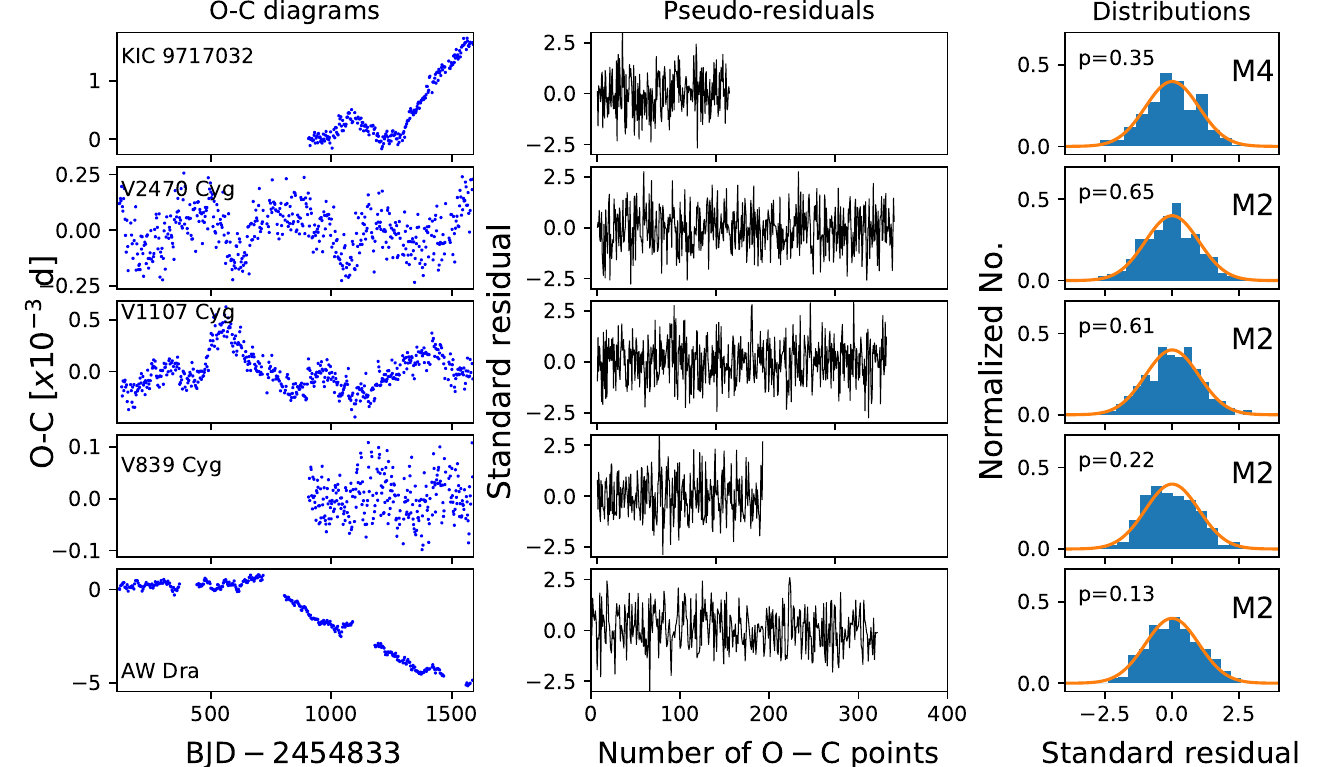}}
\caption{continued}
\label{fig:cont}
\end{figure*}

\FloatBarrier

\begin{figure*}[ht]
\section{Statistical fitting of the O$-$C curves of \textit{TESS} RRc stars.}
\subfloat{\includegraphics[width=0.8\textwidth]{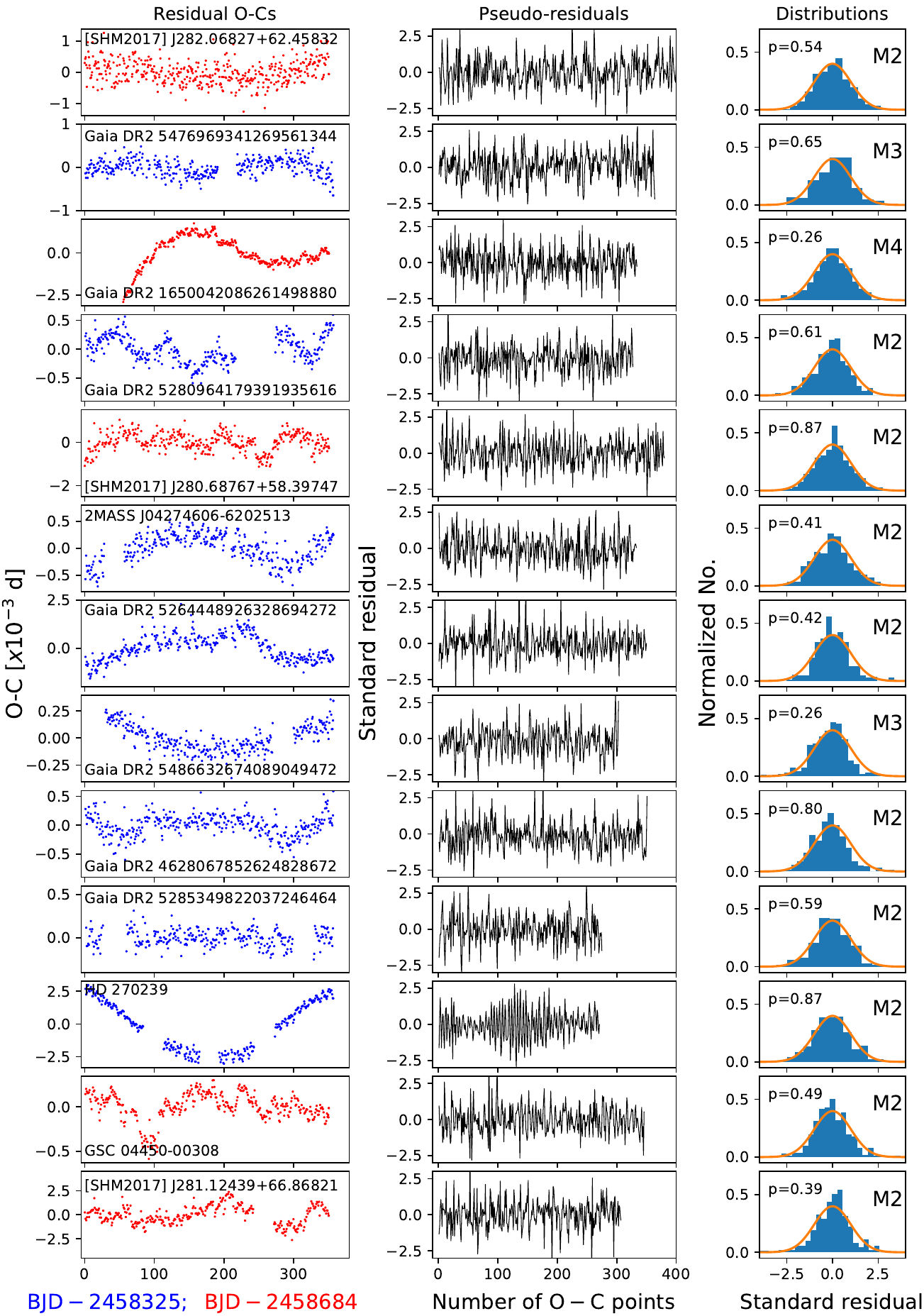}}
\caption{
First column: O$-$C diagrams calculated from the phase variation functions of \citet{Benko2023}, from which the short-period signals are pre-whitened. 
Second column: The pseudo-residuals (see formula (\ref{eq:u})) associated with the best statistical model.
Third column:  Normalised distribution of pseudo-residual values (blue histograms) compared to the standard normal distribution (orange curves). 
In the upper right corner of the panels, the resulting optimal model (M1-M4) is indicated, in the upper left corner, the p-value of the normality test is given.
}\label{fig:res_plot_rrc}
\end{figure*}
\begin{figure*}
\ContinuedFloat
\subfloat[]{
\includegraphics[width=0.8\textwidth]{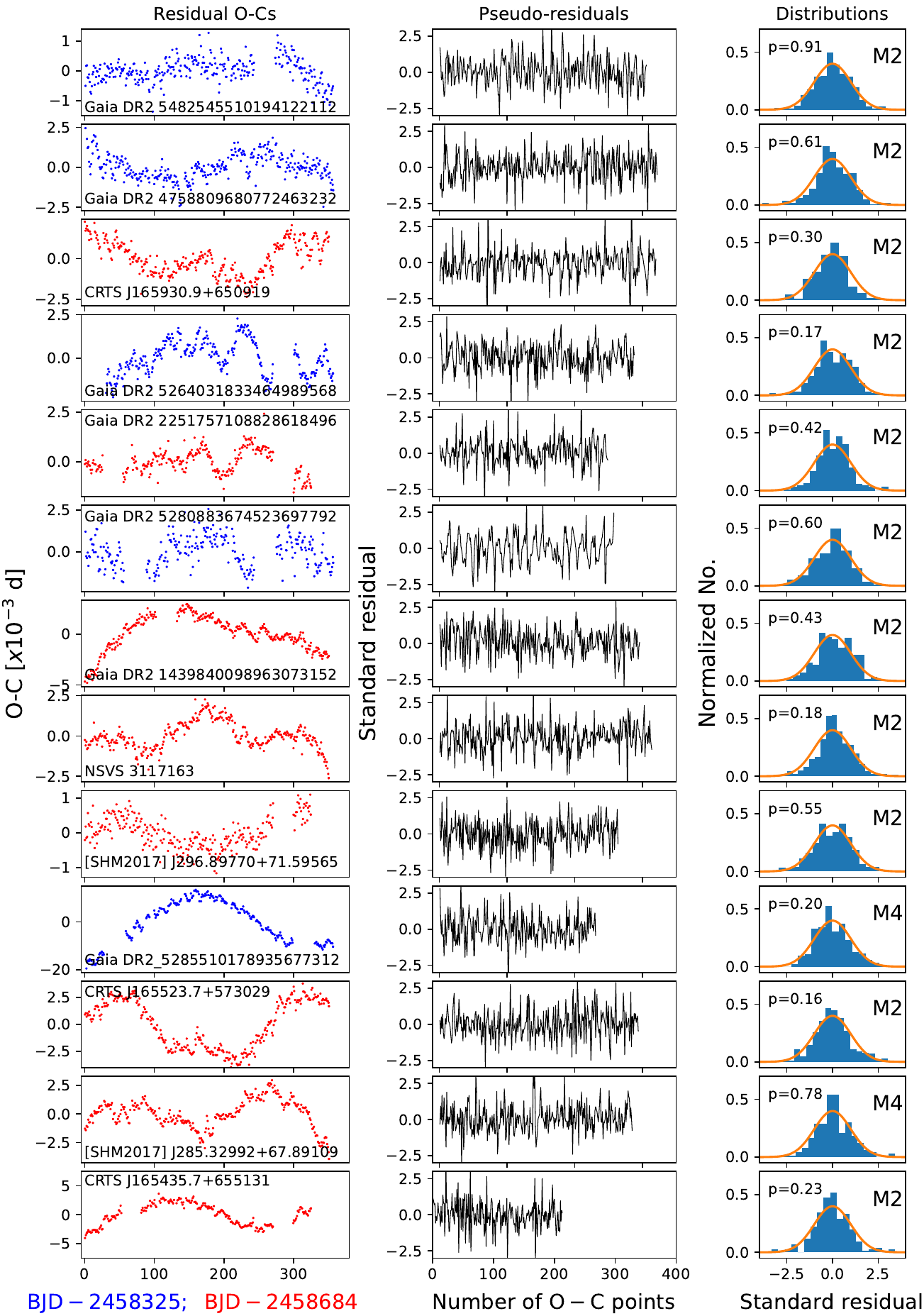}}
\caption{
continued
}\label{fig:res_2}
\end{figure*}
\FloatBarrier
\clearpage

\end{appendix}
\end{document}